# Molecular Beam Epitaxy of Wurtzite (Ga,Mn)N Films on Sapphire(0001) Showing the Ferromagnetic Behaviour at Room Temperature

Saki Sonoda[a], Saburo Shimizu[a], Takahiko Sasaki[b], Yoshiyuki Yamamoto[b] and Hidenobu Hori[b]

[a]ULVAC JAPAN, Ltd., 2500 Hagisono, Chigasaki, Kanagawa 253-8543, Japan

[b]Department of Physical Materials, School of Materials Science, Japan Advanced Institute of Science and Technology (JAIST), 1-1 Asahidai Tatsunokuchi, Nomi, Ishikawa 923-1211, Japan

Corresponding Author. e-mail address: saki_sonoda@ulvac.com   (S. Sonoda)

**Abstract**

Wurtzite (Ga,Mn)N films showing ferromagnetic behaviour at room temperature were successfully grown on sapphire(0001) substrates by molecular beam epitaxy using ammonia as nitrogen source. Magnetization measurements were carried out by a superconducting quantum interference device at the temperatures between 1.8K and 300K with magnetic field applied parallel to the film plane up to 7T. The magnetic-field dependence of magnetization of a (Ga,Mn)N film at 300K were ferromagnetic, while a GaN film showed Pauli paramagnetism like behaviour. The Curie temperatures of a (Ga,Mn)N film was estimated as 940K.

Keywords: (Ga,Mn)N; MBE; ammonia; ferromagnet; DMS

Research on diluted magnetic semiconductors (DMSs) has attracted much interest since the first fabrication of the Mn doped InAs, (In,Mn)As[1], and has led to the following intensive investigations on the  -  based DMSs[2-4] and the realization of electrical spin injection into a non-magnetic semiconductor using a ferromagnetic semiconductor (Ga,Mn)As as a spin polarizer[5]. However, the Curie temperature (Tc) of the GaAs-based DMS is as low as 110K, currently. This fact means that the function is available only at low temperatures. In the meanwhile, the recent development of growth techniques for wurtzite  -nitrides has resulted in the successful fabrication of GaN-based optical and electronic devices[6-7]. Moreover, there are some theoretical predictions about the possibility of GaN-based DMS having a Tc exceeding RT[8-9]. Therefore fabrication of magneto optical and electronic devices that operated at RT will be enabled by realization of the ferromagnetism with a high Tc in GaN-based DMS. To our knowledge, there are a few studies on the magnetic characteristics of GaN-based DMSs[10], but no reports on the ferromagnetic characteristics at temperatures exceeding RT.

The aim of this study is to realize DMS films based on GaN. In this paper, we report, for the first time, the successful growth of wurtzite (Ga,Mn)N films on sapphire(0001) substrates by molecular beam epitaxy using ammonia as nitrogen source ($NH_3$-MBE) that exhibit the ferromagnetic characteristics at RT.

The growth of (Ga,Mn)N films was carried out in an $NH_3$-MBE system (ULVAC, MBC-100) equipped with a reflection high energy electron diffraction (RHEED) apparatus. Solid source effusion cells were used as Ga and Mn sources while ammonia gas was used as the nitrogen source. (Ga,Mn)N films of thicknesses ranging from 1300Å to 3600Å were grown at growth temperatures between 580°C and 720°C and at various Ga/Mn flux ratios on wurtzite GaN(0001) buffer layers on sapphire(0001) substrates[11]. RHEED pattern observation revealed that single crystal wurtzite (Ga,Mn)N films are obtained at a higher growth temperature and at a higher Ga/Mn flux ratio. Also zinc-blende crystals were included in the films grown at a lower growth temperature and at a lower Ga/Mn flux ratio. The growth condition dependences of the crystal structures of the grown films will be discussed in detail in Ref. [12]. After the growth of the (Ga,Mn)N layer, a 200Å-thick GaN layer was grown for use as a cap layer to prevent oxidation of the (Ga,Mn)N layer. Two (Ga,Mn)N films which showed 1×1 RHEED patterns were prepared for measurements of magnetic and electrical properties. Table I gives the growth conditions and the RHEED patterns observed during the growth of these (Ga,Mn)N layers (sample A and sample B). X-ray diffraction measurements revealed that both samples had wurtzite structures without phase separation[12]. This is consistent with the observation of RHEED patterns of the (Ga,Mn)N layers. A GaN film of 3000Å thickness on a GaN buffer layer was grown at 720°C [12] for a comparison of magnetic properties.

After the growth, magnetization measurements of the wurtzite (Ga,Mn)N films, the GaN film and the sapphire(0001) substrate were carried out by a superconducting quantum interference device (SQUID) in the temperature range 1.8K to 300K and the magnetic field up to 7T applied parallel to the plane of the film. Figure 1 shows the relationship between magnetic-field H and the magnetization M (M-H curve) of sample A at 1.8K 4.2K, 8.0K, 12K and 300K. The GaN film and the Sapphire(0001) substrate showed Pauli paramagnetism like and diamagnetic behaviours, respectively, in the temperature range from 1.8K to 300K[13]. In Fig.1, the contributions of the GaN buffer layer and sapphire(0001) substrate have been subtracted from the data of sample A. The inset of Fig. 1 represents M-H curves of sample A and the GaN film at 4.2K including the contribution of the diamagnetic behaviour in the sapphire(0001) substrate. Sample B showed the same tendency in its M-H curves at 1.8K, 4.2K, 8.0K, 12K and also 300K[13]. As representative data, the coercivity Hc and the residual magnetization Mr at 300K for sample A are summarized in Table II. From these M-H curves, it can be stated that the (Ga,Mn)N films are ferromagnetic at 300K, whilst there is a possibility that some paramagnetic aspect might coexist with the ferromagnetic phase that can be seen in the anomalous increase of high field magnetization in the low temperature region. To determine a Tc of sample A, the magnetization values were measured against temperatures in the temperature range 300K to 750K at the magnetic field of 0.1T. Figure 2 shows the M-T curve of sample A. A critical point of magnetic disordering of the film could not be observed directly due to the operating-temperature limit, 750K, of components of SQUID equipment such as leading-out electrodes and a holder that a sample is fixed to. The Tc of the sample A was estimated as 940K by extrapolation from the data of the M-T curve using mean field approximation[13].

Table I. Growth conditions and RHEED patterns of the (Ga,Mn)N layers of sample A and sample B.

|  | Sample A | Sample B |
|---|---|---|
| Cell temperature (°C) | | |
| Ga | 850 | 825 |
| Mn | 575 | 575 |
| $NH_3$ flow (sccm) | 5 | 5 |
| Substrate temperature (°C) | 720 | 720 |
| Thickness (Å) | 3600 | 2000 |
| RHEED pattern | | |

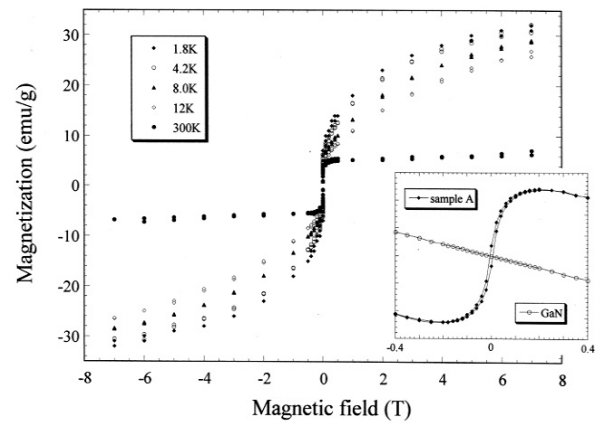

Fig. 1. Magnetic-field H dependence of the magnetization M (M-H curve) of sample A at 1.8K 4.2K, 8.0K, 12K and 300K up to 70 kOe. Inset represents M-H curves of sample A and GaN film at 4.2K including the contribution of the diamagnetic behaviour in the sapphire(0001) substrate.

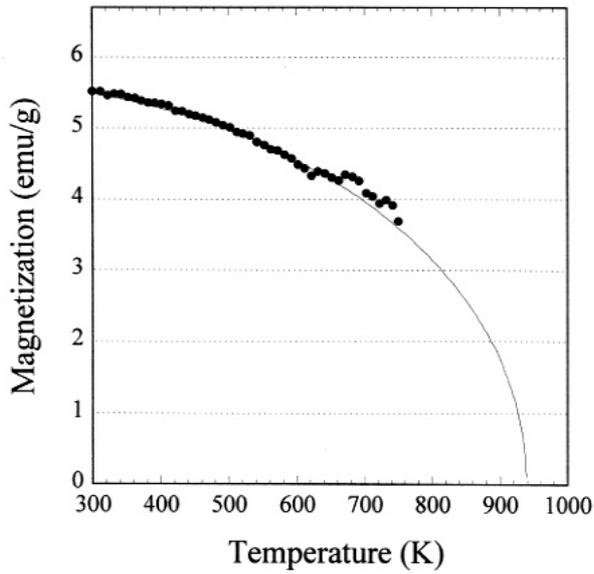

Fig. 2. Temperature T dependence of the magnetization M (M-T curve) of sample A at 0.1T.

Generation of ferrimagnetic MnGa with a Neel temperature ($T_N$) of 470K[14] or $Mn_4N$ with a $T_N$ of 740K[15] in the films is one of the possible explanations for the residual magnetization at 300K. Assuming the existence of ferrimagnetic MnGa and/or $Mn_4N$, some aspects of magnetization behaviour in the high temperature region should be observed, such as a large reduction in the magnetization value at around the $T_N$s of these ferrimagnetic materials and a negative Weiss temperature extrapolated from a plot of inversed magnetic susceptibility vs. temperature (1/

Table II. Estimated data for Magnetic Properties of sample A and sample B.

|  | Sample A | Sample B |
|---|---|---|
| Mn content (at%) | 9 | 6 |
| $\mu_{ex}$ at 300K (emu/g) | 5.5 | 0.51 |
| Hc at 300K (Oe) | 85 | 52 |
| Mr at 300K (emu/g) | 0.77 | 0.08 |
| Hc at 1.8K (Oe) | 70 | 55 |
| Mr at 1.8K (emu/g) | 1.3 | 0.3 |

Here, $\mu_{ex}$, Hc and Mr are the experimentally-determined magnetic moment, coercivity and residual magnetization, respectively.

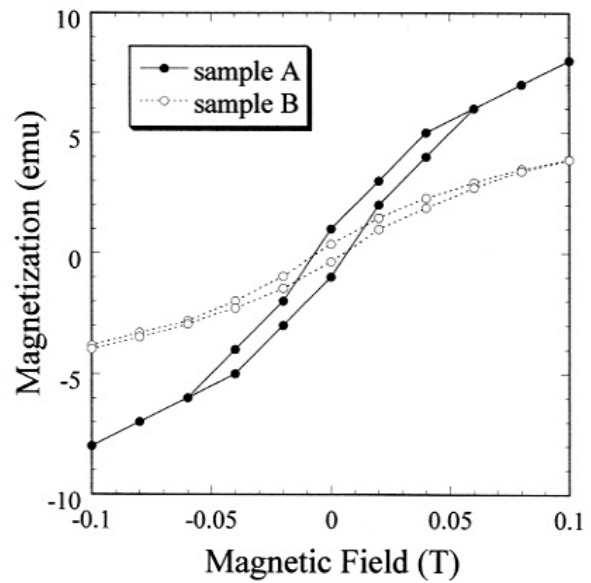

Fig 3. The hysteresis loops of sample A and sample B at around zero magnetic field at 1.8K.

-T plot). However, there is no reduction in the magnetization values at around the $T_N$s[13]. Also positive paramagnetic Curie temperatures were observed for both samples[13]. These facts negate this assumption. Also it should be emphasized that the Tc of the grown film exceeds these $T_N$s.

The magnetization curves in this study look similar to the ones shown in Ref [16] by S. Haneda et al. They discussed the possibility of existence of Fe clusters. Their model can not be directly applied to this case due to the antiferromagnetism in Mn metal. Note again that the X-ray diffraction measurements denied the existence of a Mn metal phase in both sample A and sample B. These facts mean that the appearance of ferromagnetism in the grown (Ga,Mn)N films is intrinsic as with DMS.

The experimentally-determined magnetic moments are 5.5 emu/g and 0.51 emu/g at 300K in sample A and sample B, respectively. Assuming the tri-valent state of substituted Mn in (Ga,Mn)N, the Mn concentration can be estimated from calculations using the total moment. The Mn concentrations of sample A and sample B are 9 at% and 6 at%, respectively. The hysteresis loops of sample A and sample B at around zero magnetic field at 1.8K are showed in Fig. 3. Sample A with its higher Mn content shows a larger Hc value than sample B. Table II summarizes the derived data for the magnetic properties of sample A and sample B.

With respect to the electrical properties of these samples, the carrier type was determined by Hall effect measurements to be n in both samples at the temperatures of 300K and 4.2K[13]. Due to the fact that the contributions of the cap and the buffer layers can not be easily distinguished, carrier concentrations in the (Ga,Mn)N layers have not been determined, so far. However, considering the fact that hydrogen acts as a shallow donor in GaN with p-type dopant grown in a hydrogen ambient[17], the carrier type of our as-grown (Ga,Mn)N film, is considered to be n. In this case, regarding the mixed phase problem, that is the coexistence of paramagnetic phase with ferromagnetic one, a tentative but likely model can be proposed here under following assumptions: n-type carriers or electrons participate in the stabilization of the ferromagnetic ordering through s-d exchange interaction. That is, most of Mn ions in (Ga,Mn)N are ferromagnetically coupled through the spin polarization generated by s-d exchange interaction. But some other ions are isolated at low temperature because of low electron density in this temperature region. In a high temperature region, the stability of the ferromagnetic ordering becomes prominent because of high electron density produced by thermal excitation. The model of s-d exchange interaction for ferromagnetic ordering assumed from the experimental results is in conflict with the model of p-d exchange interaction for ferromagnetic behaviour of p-type GaN with Mn[8]. This discrepancy may be due to the manipulations of the Mn valence state in the grown films that tend to have n-type conductivity due to the hydrogen incorporation and/or nitrogen vacancy that may make the crystal field in the grown films different from the one in ideal GaN film.

To clarify the detailed mechanisms of magnetic orderings in our samples, more detailed investigation is required. For that, we are preparing (Ga,Mn)N films having various Mn- and/or H-concentration grown on semi-insulating GaN buffer layers[18] and post-growth treatmants such as thermal annealing to remove hydrogen from the (Ga,Mn)N layer.

In conclusion, (Ga,Mn)N films were successfully grown on sapphire(0001) substrates by NH3-MBE. The films showed ferromagnetic behaviour at room temperature, while a GaN film showed Pauli paramagnetism like behaviour. The Curie temperatures of a (Ga,Mn)N film was estimated as 940K.